# Substrate-Induced Shifts and Screening in the Fluorescence Spectra of Supramolecular Adsorbed Organic Monolayers

James Kerfoot,[1] Vladimir V. Korolkov,[1] Anton S. Nizovtsev,[2,3] Ryan Jones,[1] Takashi Taniguchi,[4] Kenji Watanabe,[4] Igor Lesanovsky,[1] Bea Olmos,[1] Nicholas A. Besley,[2] Elena Besley[2] and Peter H. Beton.[1]

[1] School of Physics & Astronomy, University of Nottingham, Nottingham NG7 2RD, U.K.

[2] School of Chemistry, University of Nottingham, Nottingham NG7 2RD, U.K.

[3] Nikolaev Institute of Inorganic Chemistry, Siberian Branch of the Russian Academy of Sciences, Academician Lavrentiev Avenue 3, 630090, Novosibirsk, Russian Federation.

[4] National Institute for Materials Science, 1-1 Namiki, Tsukuba, Ibaraki 305-0044, Japan.

## Abstract

We have investigated the influence of the substrate on the fluorescence of adsorbed organic molecules. Monolayer films of perylene-3,4,9,10-tetracarboxylic-3,4,9,10-diimide (PTCDI), a supramolecular network formed from PTCDI and melamine, and perylene-3,4,9,10-tetracarboxylic-3,4,9,10-dianhydride (PTCDA) have been deposited on hexagonal boron nitride (hBN). The principal peaks in the fluorescence spectra of these films were red-shifted by up to 0.37 eV relative to published measurements for molecules in helium droplets. Smaller shifts (~0.03 eV) arising from interactions between neighbouring molecules are investigated by comparing the fluorescence of distinct arrangements of PTCDI, which are templated by supramolecular self-assembly and determined with molecular resolution using atomic force microscopy under ambient conditions. We compare our experimental results with red-shifts calculated using a combination of a perturbative model and density functional theory which account for, respectively, resonant and non-resonant effects of a dielectric hBN substrate. We show that the substrate gives rise to a red-shift in the fluorescence of an adsorbed molecule and also screens the interactions between neighbouring transition dipole moments; both these effects depend on the refractive index of the substrate.

## 1. Introduction

The optical properties of organic molecules in 3D crystals, thin films, and in the solution phase has been studied for many decades[1–5], but it remains difficult to predict the influence of environment on fluorescence and absorption. One area of particular interest is the coupling of transition dipole moments of neighbouring molecules resulting in the formation of H- and J-aggregates which can, respectively, suppress or enhance fluorescence with accompanying blue/red spectral shifts, and also offers the prospect of a molecular implementation of super-radiance and related quantum optical effects[4,6–10]. Recently a new approach to investigating the coupling of transition dipole moments has emerged through the study of molecules on a surface using a combination of scanning probe microscopy, which provides precise information about the relative position of neighbouring molecules, and fluorescence spectroscopy. For example Müller et al. measured differences in fluorescence for distinct monolayer phases of perylene-3,4,9,10-tetracarboxylic-3,4,9,10-dianhydride (PTCDA) on alkali halide surfaces[6,11,12], demonstrating that the position and orientation of transition dipole moments within a supramolecular array can influence the fluorescence peak energy[4]. In addition, scanning probes have been used to form molecular dimers and aggregates through probe-induced manipulation; this approach facilitates a systematic study of the dependence of resonant intermolecular interactions on molecular separation and orientation[7,8]. The characteristic energy shift which arises from dipolar coupling between neighbouring transition dipoles is typically of order 20 meV, and previous studies[7,8,11,13,14] have focussed on the effect of in-plane molecular ordering on fluorescence. However, there is a much larger shift, in the range of 50 - 400 meV, between the peaks in fluorescence of molecule in the gas-phase and the same molecule adsorbed on a substrate,[15–18] and both this effect and the role of the substrate in screening the interactions between neighbouring transition dipoles have received less attention to date.

In this paper we present a study of two perylene derivatives on the hexagonal boron nitride (hBN) surface. These molecules exhibit a large, 0.3 – 0.4 eV 'gas-surface red-shift', i.e. a shift in fluorescence peak energy of an adsorbed molecule as compared with the same molecule in the gas-phase (or helium nano droplet), and also provide a system in which supramolecular organisation can be used to distinguish smaller (~0.03 eV) fluorescence shifts due to differences in molecular in-plane organisation[19,20]. We use a combination of density functional theory and a perturbative approach to provide a unified description of both substrate-induced fluorescence shifts and dipolar screening. Specifically we highlight the importance of resonant interactions with the substrate which lead to a red-shift in the fluorescence of adsorbed molecules, and also a screening of the interactions between the transition dipole moments of neighbouring molecules. These effects can be larger than, or, in some cases, comparable to the non-resonant contributions to the red-shift which can be calculated using density functional theory. The resonant interactions are determined, in part, by the dielectric properties of the substrate, and we identify a phenomenological dependence of red-shift on refractive index by combining the measurements below with data extracted from the literature. Our model is related to solvatochromism,[21–24] and represents an analogue theory for molecules on semi-infinite dielectrics, which leads to shifts in fluorescence energy which are determined by the refractive index of the substrate.

## 2. Molecular adsorption and fluorescence

hBN is chosen as a substrate for this investigation since it provides an atomically flat and weakly interacting surface which is compatible with molecular deposition and subsequent characterisation, with molecular resolution, using atomic force microscopy (AFM) under ambient conditions[25,26]. Since hBN is an insulator the fluorescence of adsorbed molecules can be measured allowing a correlation of molecular organisation, as determined by AFM, and optical properties. We use hBN flakes with typical thicknesses of a few 10s of nanometres and lateral dimensions of a few 10s of microns, which are exfoliated onto a supporting Si/SiO$_2$ substrate. The preparation of hBN flakes, deposition of molecules and imaging protocols follow our previous work[26] and are described in the Methods section. All AFM and fluorescence measurements were acquired under ambient conditions.

To investigate the dependence of molecular placement on fluorescence we have exploited two-dimensional supramolecular assembly to form two distinct networks of the fluorophore perylene-3,4,9,10-tetracarboxylic-3,4,9,10-diimide (PTCDI), each of which is stabilised by hydrogen bonding. In the first arrangement PTCDI forms a honeycomb network stabilised by hydrogen bonding with melamine[19]. This network is deposited from solution[27] and can be converted into a denser, row-like phase[20,28] of PTCDI by removal of melamine through rinsing of the PTCDI-melamine network with water. We have also investigated the fluorescence of PTCDA, which is deposited by immersion in an ethanolic solution. Schematics of these molecules are shown in Figure 1a.

The morphology and molecular arrangement of these networks are determined using AFM. Figure 1b shows AFM images of the supramolecular network formed by melamine and PTCDI following deposition from solution (see Methods). Each melamine is hydrogen-bonded to three PTCDI molecules and the three-fold rotational symmetry of melamine gives rise to an extended honeycomb network as shown schematically in Fig. 1c. The deposition of PTCDI-melamine on hBN has been reported previously[29], but in the present study the imaging and preparation protocols have been improved to allow much clearer identification of the supramolecular arrangement and the formation of larger islands with lower defect densities. The network has a lattice constant of 3.5 ± 0.1 nm, similar to arrays reported previously[19,28,29] on Ag/Si(111), Au(111), graphite and $MoS_2$.

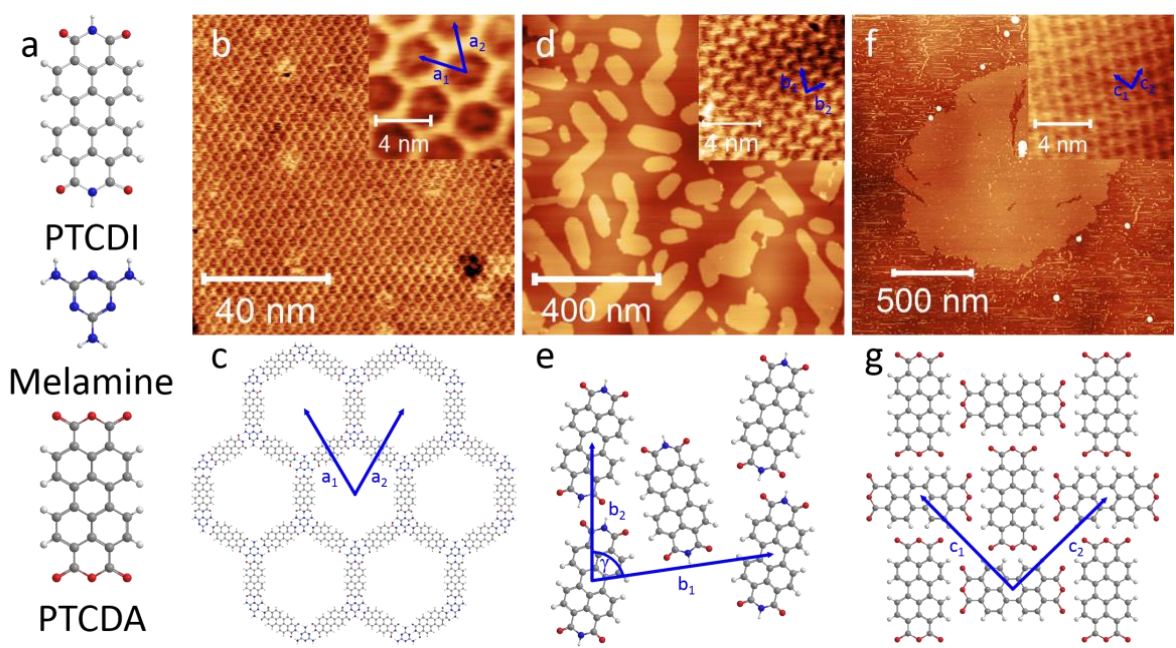

**Figure 1.** PTCDI, PTCDA and the PTCDI-melamine supramolecular network were deposited on hBN from solution: a) schematics of molecular structures; b) AFM image of PTCDI-melamine structure with inset showing honeycomb supramolecular organisation from which structural model in c) is determined; d) AFM of PTCDI with inset showing the molecular arrangement in the canted phase as shown schematically in e); AFM of PTCDA island with high resolution image of molecular arrangement in the square phase shown schematically in f). From high resolution AFM images, the following lattice constants, labelled in both AFM images and schematic diagrams, were extracted; $a_1 = a_2 = 3.5 \pm 0.1$ nm, $b_1 = 1.75 \pm 0.1$ nm, $b_2 = 1.45 \pm 0.1$ nm, $\gamma = 84 \pm 1$ ° and $c_1 = c_2 = 1.6 \pm 0.1$ nm.

Immersion of the PTCDI-melamine array in water leads to the removal of the more soluble melamine and converts the network into islands of PTCDI with monolayer height. AFM images of PTCDI islands including high resolution scans (see Fig. 1d), show that the PTCDI molecules are arranged in rows, with inter-row ($b_1$) and intra-row ($b_2$) separations of 1.75± 0.1 nm and 1.45 ± 0.1 nm respectively, and an angle $\gamma$ = 84 ± 1° between lattice vectors. These parameters are in good agreement with previous investigations using scanning tunnelling microscopy (STM) on graphite[30], Ag/Si(111)[20] and Au(111)[31,32]. This agreement, together with the canting of molecules relative to the row direction, which has also been observed in STM studies, provides strong evidence for head-to-tail hydrogen bonding between neighbouring molecules. Overall our images are consistent with a structural model of PTCDI monolayers which consists of parallel rows of canted molecules as illustrated schematically in Fig. 1e.

Figure 2 shows the normalised fluorescence spectra of PTCDI and the PTCDI-melamine network. Measurements were taken using a Horiba LabRam HR spectrometer with an excitation wavelength of 532 nm and a spot size of approximately 1 µm² (see Methods). The fluorescence spectra show an intense zero-phonon peak and, at lower energy, associated vibronic peaks. There is a clear difference in the energies of these peaks for different molecular arrangements; the zero-phonon peak of solution-deposited PTCDI on hBN appears at 2.214 ± 0.002 eV, which is red-shifted from the equivalent peak of the PTCDI-melamine array, which occurs at 2.245 ± 0.002 eV, by 31 ± 3 meV. These values are very close to the measured absorption peak for alkylated PTCDI derivatives adsorbed on graphene[33].

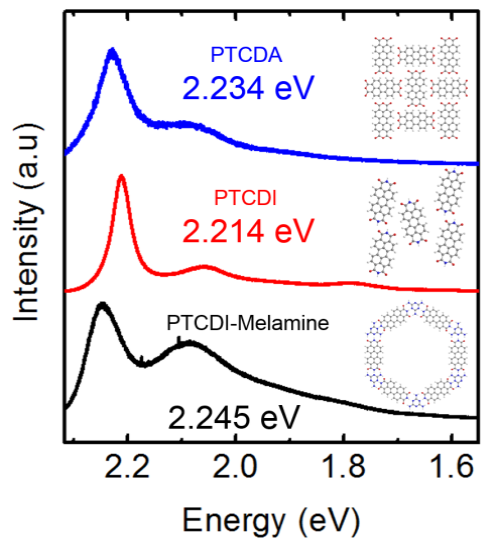

**Figure 2.** Normalised fluorescence spectra of PTCDI, a PTCDI-melamine network and PTCDA on hBN acquired with an excitation wavelength of 532 nm.

As discussed above we are also interested in a 'gas-surface' shift for these molecules; this is analogous to the 'gas-crystal' shift[23,34] which has been widely discussed for organic semiconductors and refers to changes in absorption/emission energies in the solid state as compared with the gas phase. Although PTCDI provides a suitable molecular system for the comparison of in-plane ordering, the fluorescence energy of PTCDI in the gas phase is not available in the literature. The absorption energy for PTCDI-Me (a perylene derivative in which the hydrogen of the imide group is replaced by a methyl group) has been measured[18] for a molecule adsorbed on a helium nano droplet (HND) and found to be 2.55 eV. This allows a rough estimate of the gas-surface red-shift for PTCDI of ~0.3 eV, approximately one order of magnitude greater than the differences which arise from changes in in-plane ordering.

The absence of gas-phase data for PTCDI has motivated a parallel study of PTCDA, a closely-related molecule which has been studied much more widely, including on several different substrates and on helium droplets[16,17,35]. We have prepared monolayer-thick islands of PTCDA by deposition on hBN from solution (see Methods). AFM images (Fig. 1f) show that large PTCDA islands are formed and high resolution images (Fig. 1f inset) reveal a molecular packing with square symmetry. From the observed lattice constants (1.6 ± 0.1 nm) and symmetry, the molecular organisation in this phase is consistent with that shown in Fig.1g; here alternate molecules are rotated by 90°. This phase is similar to monolayer arrangements observed on Ag/Si(111)[20,36] and other surfaces[36,37] for which a square arrangement with a lattice constant of 1.63 nm has been reported[20,36].

The fluorescence spectrum for this PTCDA phase has been measured and the zero-phonon peak is observed at an energy 2.234 ± 0.002 eV (Fig. 2). The fluorescence energy of PTCDA embedded in He droplets (which typically differ from the gas-phase value by less than 10 meV[38]) has been reported[16] to be 2.602 eV giving a red-shift $\Delta E_{PTCDA}$ = 0.368 ± 0.002 eV when the molecules are adsorbed on hBN.

**3. Substrate-induced red-shifts**

To understand the shifts in fluorescence energy we consider the interactions between a molecule (PTCDA or PTCDI) with its molecular neighbours and, also, with the underlying dielectric substrate. We first discuss the changes arising from the interaction with the substrate since these are, experimentally, larger by an order of magnitude. There are several possible contributions to the substrate-induced shift of fluorescence energy which may be usefully classified, within a perturbative approach, as resonant and non-resonant contributions[1,2,6,13]. Resonant interactions arise, in general, from the coupling of the transition dipole moment of a molecule with its environment, which in this case would include the dielectric substrate and neighbouring molecules. The non-resonant interactions arise from shifts in molecular energy levels due to surface adsorption and, in principle, can be calculated using density functional theory (DFT).

*3.1 Non-resonant effects*

Non-resonant interactions induce direct shifts in molecular energy levels due to surface adsorption. These could result from a change in molecular conformation, for example arising from van der Waals interactions[25] with the substrate, through the presence of permanent dipoles, or other mechanisms. We have calculated these effects using DFT and focus initially on the results for PTCDA. Full details of the methodology and results are provided in the SI. To summarise, the molecular geometry of PTCDA adsorbed on hBN, and also in the gas phase, were optimized using the range-separated hybrid $\omega$B97X-D functional including an empirical dispersion correction[39] in combination with the correlation-consistent cc-pVDZ basis set[40]. The hBN surface was modelled as a monolayer flake consisting of 65 boron atoms and 65 nitrogen atoms with edges terminated by H atoms. Atomic positions of the surface were initially optimized and were frozen in the subsequent calculations. Molecular adsorption energies were determined and for each molecule the most energetically preferred adsorption site was used to calculate excited state ($S_1$) geometries and related properties. Excitation energies corresponding to optical absorption ($S_1 \leftarrow S_0$) and fluorescence ($S_0 \leftarrow S_1$) were determined with the time-dependent density functional theory (TD-DFT) using the optimized structures of the $S_0$ and $S_1$ states, respectively. All calculations were performed with the Q-Chem software package[41].

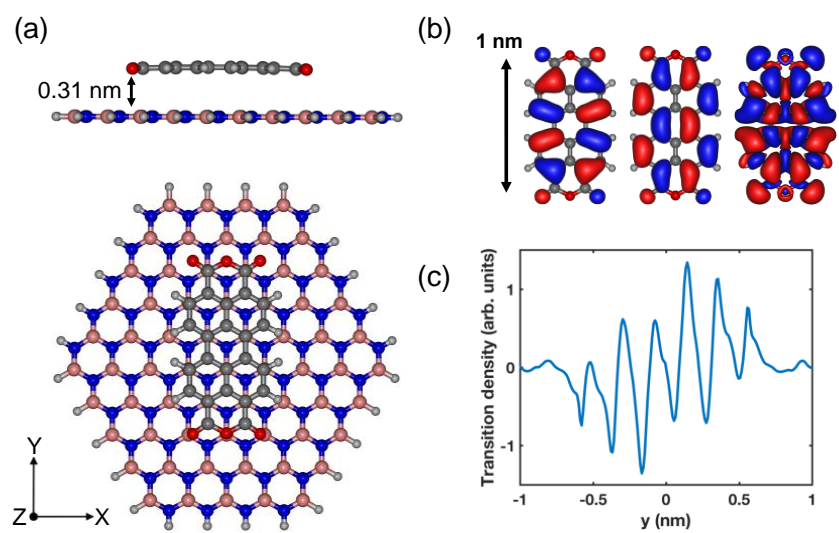

**Figure 3** Summary of results from density functional theory. a) schematics of relaxed PTCDA on hBN in cross-section (upper) and top view (lower); b) calculated probability amplitudes for the HOMO (left) and LUMO (centre) orbitals, and the transition density (right); c) reduced transition density *g(y)*.

Figure 3 and Table 1 summarise the results for the fluorescence of PTCDA; all other data appear in SI. For calculations of this type the absolute values of transition energies are in reasonably good agreement with the experimental values discussed above. We are particularly interested in the red-shift due to adsorption on the hBN surface, for which the calculated value is 0.10 eV (the difference in transition energy in Table 1 for a molecule on and off the surface). This represents the non-resonant contribution to the overall red-shift.

The presence of the dielectric hBN substrate leads to two different effects that are responsible for the shift: (i) a reduction of the HOMO (highest occupied molecular orbital)-LUMO (lowest unoccupied molecular orbital) gap of an adsorbed molecule[42,43] and (ii) a weakening of the electron-hole interaction[44]. The HOMO-LUMO gap is much bigger than the $S_0 \leftarrow S_1$ transition energy (Table 1), which can be accurately predicted using TD-DFT. However, TD-DFT with standard functionals may underestimate the shift of energy levels upon molecular adsorption[45] owing to an inaccurate treatment of the substrate polarisation effect arising from the neglect of nonlocal electron correlation effects.[46,47] Further results show that, as expected for a planar, highly symmetric molecule such as PTCDA, shifts due to the presence of a permanent dipole maybe neglected. In addition, there is a negligible distortion of the molecule on adsorption on hBN (see SI).

|  | Gas Phase | Adsorbed on hBN |
|---|---|---|
| Adsorption energy (eV) | - | 2.46 |
| Molecule-substrate separation, $d$ (nm) | - | 0.31 |
| Transition dipole moment (Debye) | (0, 8.7, 0) | (0, 7.9, 0) |
| Transition energy (eV) | 2.43 | 2.33 |
| HOMO (eV) | -8.00 | -7.81 |
| LUMO (eV) | -2.83 | -2.69 |
| HOMO-LUMO gap (eV) | 5.17 | 5.12 |

**Table 1** Calculated parameters for PTCDA adsorbed on hBN. Values are calculated for PTCDA in the excited $S_1$ state.

*3.2 Resonant effects*

The resonant interaction with the substrate may be modelled by treating the adsorbed molecule as an oscillating transition dipole moment with magnitude $\mu$ placed at a height $d$ above the hBN surface. For the case of the planar PTCDI and PTCDA molecules our calculations (Table 1) show that $d$ is in the range 0.30 - 0.35 nm and the transition dipole moment is oriented parallel to the hBN surface. Within a simple electrostatic picture a charge $q$ placed close to the interface between a region with dielectric constant $\approx 1$ and a semi-infinite dielectric with dielectric constant $\varepsilon$ induces a polarisation equivalent to the field from an image charge $q' = -q(\varepsilon-1)/(\varepsilon+1)$ placed below and equidistant from the image plane. Here we consider the image plane to be midway between the adsorbed molecule and the substrate surface (see[42,43,48,49] for a discussion of the placement of the image plane); thus there is a separation of $d/2$ between the image plane and both the real and image charges. Similarly, a dipole with moment $\mu$ induces an image dipole with moment $\mu' = -\mu(\varepsilon-1)/(\varepsilon+1)$ at a distance $d/2$ below the image plane. In the subsequent discussion we replace the relative permittivity, $\varepsilon$, with $n^2$ where $n$ is the refractive index of the substrate.

These electrostatic effects may be incorporated into a quantum mechanical calculation of a two level system with transition dipole moment $\mu$, placed close to the interface of a dielectric by considering the interaction between the real and image dipoles. As we show in detail in Supplementary Information (SI), this leads to a red-shift $\Delta E_{subs}$, of the emission energy which is given by

$$\Delta E_{subs} = \frac{n^2-1}{n^2+1}\frac{\mu^2}{4\pi\varepsilon_o d^3}. \tag{1}$$

This result is valid in the limit $d \ll \lambda$, the wavelength of the emitted light and is derived by considering the perturbative effect of a dielectric environment, and can also be evaluated within a Green's function formalism[50] (see SI - in this approach the image charges are not treated explicitly; instead the formalism ensures that the electrostatic boundary conditions at the dielectric interface are satisfied).

Classically this energy may be identified as the dipolar coupling between the transition dipole and a dipole with the magnitude of its image (as discussed below an additional factor of ½ appears in a simple calculation of the potential energy of a dipole due to its image; this term, with the additional factor of 1/2, has been used previously to estimate the solvatochromic shift, an analogous theory which we discuss below). The energy given by equation (1) corresponds to a resonant red-shift which results when a molecule is transferred from the gas phase to an adsorbed state on the substrate, and is expected in addition to any (non-resonant) shifts calculated using DFT.

According to equation (1) we should expect a clear dependence of energy shift on the refractive index of the substrate. In Figure 4 the dependence of the red-shift on the refractive index is confirmed; here we have extracted from the literature the zero-phonon peak position of PTCDA adsorbed on various alkali halides and mica[6,11,51,52] together with our results above for PTCDA on hBN (an additional point for sublimed PTCDA on hBN is also included; see SI for more details). These are converted to a value for substrate-induced red-shift using the peak position (2.602 eV) measured[16] for PTCDA on a He droplet (see above). These values are then plotted versus $(n^2-1)/(n^2+1)$ using values for the refractive index of each substrate which are included in Fig. 4 inset. Note that hBN is a negative uniaxial material with ordinary and extraordinary refractive indices[53] $n_o$ = 2.13 and $n_e$ = 1.65 respectively. For a uniaxial dielectric the effective permittivity determining the image charges (see above) is a geometric average[54] of the diagonal components of the dielectric tensor and the effective refractive index satisfies $n^2 = n_o n_e$, which gives $n = 1.87$.

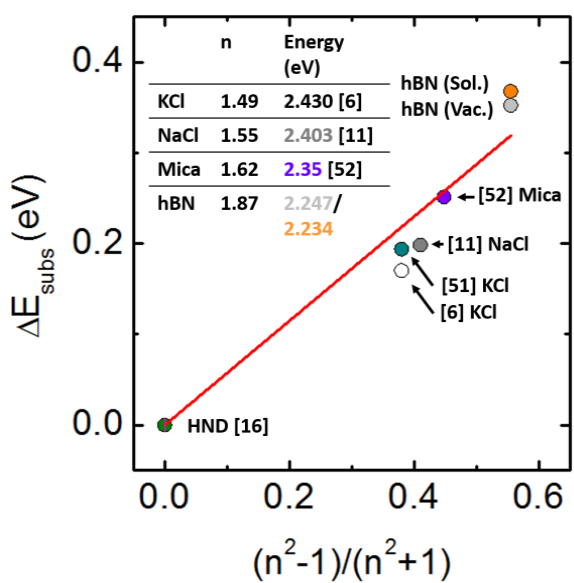

**Figure 4** The shift of the fluorescence peak of PTCDA adsorbed on various surfaces plotted against the predicted dependence of the shift on refractive index according to equation (1). The results for fluorescence on mica and alkali halides are extracted from the literature (references in square brackets) and the measured values for the peak energy, and the refractive index of the substrate are included in a Table in the inset. Peak energies measured for PTCDA on hBN deposited by sublimation and from solution are also included. The reference energy is the value measured for PTCDA on a helium nano droplet (HND)[16]. Values of peak energies are derived from the fluorescence of extended supramolecular arrays.

Figure 4 reveals a systematic increase in red-shift which increases for substrates with larger refractive index with a functional dependence which is in reasonable agreement with the form predicted by equation (1); a straight line fit to the data gives a gradient of 0.572 eV. Note that the data points included in Fig. 4 are measured for extended two-dimensional layers of molecules rather than isolated molecules on the surface. The shift in Fig. 4 therefore includes the contribution from both the substrate and the in-plane shift due to the presence of nearest neighbours. However, as discussed above, the substrate-induced shift is larger, typically by an order of magnitude, than the in-plane shifts. The results in Fig. 4 confirm, phenomenologically, that the dominant contribution to the overall red-shift is related to the refractive index of the substrate.

The data in Table 1 (calculated transition dipole moment and molecular-substrate sepearation) give a predicted value for $\mu^2/4\pi\varepsilon_o d^3 = 1.3$ eV, which when combined with refractive index dependent term in equation (1) gives a resonant shift for PTCDA of 0.72 eV. Combined with the non-resonant shift discussed above this results in an overall calculated shift of 0.82 eV which is significantly greater than the observed shift of 0.368 eV.

To understand the origin of this difference we re-visit one of the key assumptions in the simple theory above, which is that the transition dipole moment may be treated as a point source. In fact, it arises from variations in charge density which are distributed over the molecule; thus the transition dipole moment has a finite size comparable with the molecular dimension, $l$. The assumption of a point dipole is valid only if the characteristic separation, $d$, between the dielectric and molecule satisfies $d > l$, but for a large planar molecule such as PTCDA, $l > d$, so this assumption does not hold. We include a heuristic correction[23,55–57] to the energy, the extended dipole model, by assuming that the dipole can be represented as two charges $\pm \mu/\delta$ separated by a distance $\delta$ positioned at a height $d$ above a dielectric surface. Classically this leads to a reduction in the electrostatic energy given in equation (1) by a factor $f(\delta/d) = 2(d/\delta)^2(1-(1+(\delta/d)^2)^{-1/2})$.

The parameter $\delta$, which characterises the charge separation can be estimated from our DFT results. The transition dipole moment for emission is calculated from the electron wave functions $\varphi_{LUMO}$ and $\varphi_{HOMO}$ of, respectively, the initial (LUMO) and final (HOMO) states as follows,

$$\boldsymbol{\mu} = e \int_{-\infty}^{\infty} \varphi_{LUMO}^*(x,y,z)\, \boldsymbol{r}\, \varphi_{HOMO}(x,y,z)\, d\boldsymbol{r},$$ where $e$ is the electronic charge. These wavefunctions, and their product which appears in the integrand, are shown schematically in Fig. 3. Due to the symmetry of PTCDA the dipole moment is oriented along the y-axis (see Fig. 3) and the above integral reduces to $\mu = \int_{-\infty}^{\infty} y\, g(y)\, dy$, where

$$g(y) = e \int_{-\infty}^{\infty} \varphi_{LUMO}^*(x,y,z)\, \varphi_{HOMO}(x,y,z)\, dx\, dz$$

This function, the transition density, is plotted in Fig. 3c and corresponds to the spatial variation of the charge density associated with the oscillating transition dipole. We estimate the parameter $\delta$ as the difference between the average separation of positive and negative charge, $\delta = \mu / \int_0^{\infty} g(y)\, dy$.

A numerical calculation, based on $g(y)$ derived from our DFT results gives $\delta$ = 0.86 nm ($\delta/d$ = 2.7), a reduction factor $f(2.7) = 0.18$, and a resonant shift of 0.13 eV; this gives a predicted overall red-shift due to adsorption of 0.23 eV, which is closer to the observed value but still shows a significant deviation, and the possible reasons for this difference are discussed below.

**4. Substrate-induced screening of intermolecular interactions**

The red-shift discussed for PTCDA is calculated for an isolated molecule adsorbed on the substrate. However, as discussed above, and by several other groups[6–8,13,23], additional red-shifts occur due to coupling of the transition dipoles of neighbouring molecules. It is not possible to explore this type of red-shift through a systematic study of the adsorption of PTCDA on hBN under the experimental conditions used here, since only one phase of PTCDA is formed. However, we exploit the distinct in-plane molecular arrangements available through supramolecular organisation of the closely-related PTCDI molecule to investigate this relatively small red-shift.

Classically, the electrostatic field experienced by a second transition dipole moment which lies in the same plane and is separated by a distance $a$ is reduced due to the presence of an image dipole; for $a \gg d$, a condition which is satisfied for these molecular arrangements, the field appears to arise from a dipole with an effective magnitude $\mu_{\text{eff}} = (\mu+\mu')$, the sum of a neighbouring (real) dipole and its image, giving $\mu_{\text{eff}} = 2\mu/(\varepsilon+1)$. Accordingly the dipolar interaction between two (real) dipoles on the surface is reduced by a screening factor $2/(\varepsilon+1)$ (this factor is also the inverse of the effective dielectric constant for a charge placed at the interface between free space and a dielectric with relative permittivity $\varepsilon$). The quantum mechanical calculation discussed above can be extended to consider the resonant interaction between neighbouring molecules on a surface; these are modelled as a pair of two level systems, each with transition dipole moment, $\mu$, placed close to the interface of a dielectric. A complete discussion of this calculation is presented in SI and confirms that the screening factor which is derived using the simple classical argument above, is correctly reproduced by a full quantum mechanical analysis in the limit $a \ll \lambda$.

Our approach to the calculation of red shifts follows Sokolowski and co-workers[6] who derived the excitonic band structure which results from the interaction between neighbouring transition dipoles. The band structure is calculated using the tight binding model introduced by Davydov[2] and we modify this approach by replacing the unscreened dipolar interaction used in previous work by the screened interaction. The calculated exciton band structures for PTCDI and PTCDI-melamine, showing the energy of the delocalised excitons a function of two-dimensional wave-vector, $k$, are shown in SI; the Brillouin zones for each supramolecular arrangement, and the relative positions of neighbouring molecules, are determined from AFM images such as shown in Fig. 1. For both arrangements of PTCDI the exciton dispersion has a minimum at $k = 0$, the $\Gamma$ point of the Brillouin zone where the energy is given by,

$$\Delta E_0 = \chi_{00}^2 \frac{2}{1+n^2} \frac{1}{4\pi\varepsilon_o} \sum_{j=1}^{neighbours} \left( \frac{\boldsymbol{\mu}_i \cdot \boldsymbol{\mu}_j}{|\boldsymbol{r}_{ij}|^3} - 3\frac{(\boldsymbol{\mu}_i \cdot \boldsymbol{r})(\boldsymbol{\mu}_j \cdot \boldsymbol{r})}{|\boldsymbol{r}_{ij}|^5} \right) \quad (2)$$

which is simply the sum of the screened dipolar interactions between a molecule at position $\boldsymbol{R}_i$ and all other molecules (at sites $\boldsymbol{R}_j$) within the supramolecular array ($\boldsymbol{r}_{ij}$ is the displacement vector $\boldsymbol{R}_i - \boldsymbol{R}_j$). A negative value corresponds to a red-shift of a molecule within the array as compared with an isolated adsorbed molecule. The additional factor, $\chi_{00}^2$, in equation (2) is the Franck-Condon factor which appears in the Davydov formalism to account for vibronic effects.

The differences between the minimum energies of the PTCDI and PTCDI-melamine networks contribute to the experimentally observed shift in peak position shown in Fig. 2. The transition dipole moment and adsorption energy of PTCDI have been calculated using the DFT methodology described above (see Table 2; further details are included in SI).

|  | Gas Phase | Adsorbed on hBN |
| --- | --- | --- |
| Adsorption energy (eV) | - | 2.52 |
| Molecule-substrate separation, $d$ (nm) | - | 0.31 |
| Transition dipole moment (Debye) | (0, 8.8, 0) | (0, 8.0, 0.1) |
| Transition energy (eV) | 2.41 | 2.31 |
| HOMO (eV) | -7.65 | -7.54 |
| LUMO (eV) | -2.50 | -2.42 |
| HOMO-LUMO gap (eV) | 5.15 | 5.12 |

**Table 2** Calculated parameters for PTCDI adsorbed on hBN. Values are calculated for PTCDI in the excited $S_1$ state.

We have also calculated various transition energies to determine the influence of non-resonant effects; specifically we have calculated the fluorescence energy of a PTCDI molecule hydrogen-bonded to (i) two melamine molecules and (ii) two naphthalene tetracarboxylic di-imide (NTCDI) molecules to mimic the H-bonding in the canted phase of PTCDI. In both cases the calculation was performed in the gas phase and we find that the hydrogen bonding leads to changes in the transition dipole moment: for an isolated PTCDI molecule in the gas phase $\mu$ = 8.84 D which increases to 10.3 D and 10.1 D for PTCDI(NTCDI)$_2$ and PTCDI(melamine)$_2$ respectively. The calculated permanent dipole of gas-phase PTCDI is less than 0.1 D and may be neglected. Full details of these calculations are provided in the SI.

The parameters above may be combined with the measured geometric arrangement of PTCDI molecules in each phase to determine the difference in exciton energies for PTCDI and PTCDI-melamine. For a quantitative estimate we use a value for the Franck-Condon factor, $\chi_{00}^2$ = 0.73, taken from the Huang-Rhys factor in Megow et al.[23], which in turn is based on experimental data for PTCDI in solution[58]. The principal source of errors in Table 3 is the uncertainty in the geometrical parameters measured using AFM.

|  | Unscreened (meV) | Screened (meV) |
| --- | --- | --- |
| PTCDI | 131 ± 23 | 59 ± 10 |
| PTCDI-melamine | 64 ± 5 | 28 ± 2 |
| Relative shift | 67 ± 24 | 31 ± 10 |

**Table 3.** The calculated resonant shifts due to unscreened and screened in-plane coupling of transition dipole moments derived from band structure calculations in SI for both PTCDI and PTCDI-melamine. The relative shift between solution deposited PTCDI and PTCDI-melamine is also shown (lowest row) and is a difference between the shofts in the two different arrangements.

The calculated screened relative shift in Table 3 matches the measured difference, 31 meV, between the principal peaks of the fluorescence spectra of PTCDI and PTCDI-melamine arrays (see above).

There are several additional non-resonant effects which might contribute to the shift. These include, for example, differences in the transition energy of PTCDI due to the two different H-bonded configurations and variations of the alignment and placement of PTCDI molecules relative to the substrate. DFT calculations can be used to estimate these shifts but the calculated values in each case are typically 10 meV or less, and we omit these contributions since they are comparable to both the finite size effects related to the hBN cluster which is used to model the substrate, and also the estimated experimental error. Accordingly, we note that when screening is included the calculated resonant shift is significantly closer to the experimentally observed value supporting our argument that screening is relevant to intermolecular coupling. Our confidence in the parameters which we have used is discussed in more detail in the subsequent section.

## 5. Discussion

Our theoretical model shows that the refractive index of the substrate is expected to strongly influence the transition energies which determine the fluorescence spectrum of adsorbed molecules. Our experimental data are consistent with the predicted trends, namely that the substrate-induced red-shift monotonically increases with refractive index and is consistent with the expected proportional dependence on $(n^2-1)/(n^2+1)$. Furthermore, the screening factor for in-plane coupling is consistent with our observations within the constraints of the precision of the relevant parameters (see discussion below).

The dependence on refractive index is suggestive of analogues with the solvatochromic effect which accounts for shifts in fluorescence transition energies through the interaction of the transition dipole moment of a solvated molecule with the image dipole induced in the surrounding dielectric medium (the solvent)[22,59–62]. In the original paper by Bayliss[22] the red-shift, $\Delta E_{solv}$, was assumed to be equal to the classical energy of interaction, $-\mu E_R/2$, where $E_R$ is the 'reaction' field due to the image charge, giving $\Delta E_{solv} \propto (n^2-1)/(2n^2+1)$ where the constant of proportionality is determined by the size of the solute molecule which is assumed to occupy a spherical solvent-free cavity. The factor ½ in the classical energy arises since the charges are interacting with their image charges. Interestingly, in the quantum mechanical treatment of a two-level system the factor ½ is absent (in our case the reaction field arises from the image dipole). Within the theory of solvatochromism, the solvent also provides a screening factor reducing the apparent magnitude of the transition dipole moment by a refractive index-dependent factor[3,62], $3n^2/(n^2+2)$. The forms for the substrate-induced red-shift and the screening factor, $2/(n^2+1)$, which we derive for an adsorbed molecule differ from those for a molecule immersed in a solvent, but the effects are closely. It would therefore be expected, in line with our observations, that a substrate analogue of the solvatochromic effect would be expected to influence the fluorescence of adsorbed molecules.

While the general trends which we observe are consistent with the predictions of our simple model, the specific magnitude of the effect is difficult to predict due to limitations in the calculations and knowledge of the relevant parameters. The quantum mechanical calculation assumes the substrate can be treated as a continuum dielectric, although on the length scales of interest this must represent an approximation. Furthermore, the placement of the image plane midway between the molecular and hBN planes of atoms also represents an approximation and neglects the atomic granularity of the adsorbate/substrate system on the relevant length scale. We note that Forker et al.[14] also consider the interaction between an adsorbed organic molecule and a substrate by considering interactions with an image dipole. In their case the molecule is adsorbed on a hBN monolayer grown on a metal substrate although the hBN is treated as an electrostatically inert vacuum-like layer and the image charge is assumed to be located in the underlying metal; this leads to a much higher assumed value of separation of real and image charges and a negligible predicted substrate-induced shift. The most significant uncertainty in our quantitative prediction of the substrate-induced resonant shift relates to the treatment of the transition dipole as either a point or extended object (other models have been considered for closely-spaced fluorophores[63,64]). A rigorous treatment for this problem is not currently available and a small change in either the estimated reduction factor, or the position of the image plane might lead to much better agreement between our model and the red-shift of PTCDA.

While our data show a systematic increase in red-shift for substrates with progressively higher refractive index it is interesting to note that, according to our calculations the resonant shift accounts for only ~ 60% of the overall shift. This implies that, phenomenologically, the non-resonant shift should also increase with refractive index. As discussed above the presence of the hBN electron system is expected to to lead to screening of the Coulombic intramolecular intereactions which are, at least partially, captured in DFT calculations. We note that the problem of substrate-induced screening of excitons currently attracts much attention in the 2D materials community[65] where pronounced photoluminescence peak shifts are observed between freestanding and supported monolayers of, for example, $WS_2$. These shifts have been attributed to substrate-induced screening of the electron-hole interaction which, similar to our observations, depends on the refractive index of the substrate.

The quantitative estimates discussed above also rely on the values of transition dipole moment and the non-resonant shifts calculated using TD-DFT. The calculated value of transition dipole moment for PTCDA, 8.7 Debye, in the gas phase is reasonably close to the value inferred experimentally, 7.4 ± 0.7 Debye, by Hoffmann et al.[3] providing confidence in our calculations. The limitations of TD-DFT methodology (the choice of structural model, functional, basis set etc.) can lead to uncertainty in transition energies which are typically of the order of 20 meV. Although there is very good agreement between the observed red-shift due to in-plane ordering of PTCDI and the calculated screened resonant shift, we stress that this result depends strongly on the value of transition dipole moment (and its dependence on the hydrogen bonding interactions with neighbouring molecules) and the Franck-Condon factor. These parameters have also been considered more extensively for PTCDA and it has been shown[66] that the Franck-Condon factor is reduced when PTCDA is adsorbed on a substrate (KCl), a possible effect which is not considered here.

One effect which we have not considered in our discussion is the alignment of the valence and conduction bands of hBN with the HOMO and LUMO of the molecule. The optical band-gap of hBN is reported[67,68] to be 5.9 eV and the electron affinity has been reported[69] to be ~1.0 eV. Thus, the calculated LUMOs (see Tables 1 and 2 above) lie within the hBN gap, but the HOMOs lie ~1 eV below the valence band of hBN. For a semiconductor heterojunction, and treating the HOMO as analogous to a valence band, we might expect a hole to be transferred from the molecular HOMO to the hBN valence band. However, this simple model does not take account of the large excitonic shift which reduces the HOMO-LUMO gap from the calculated value of ~5 eV to the observed (and calculated) transition energy which is ~2.2-2.5 eV. A transfer of a hole from the molecule to the HOMO would result in the formation of an indirect exciton which would be expected to have significantly lower binding energy. It is likely therefore that this reduction in excitonic energy would make hole transfer energetically unfavourable. Nevertheless, this question merits further investigation using a more rigorous theoretical approach.

## 6. Conclusions

The high resolution which can be attained using AFM under ambient conditions allows the identification of molecular arrangements with a precision that allows the estimation of the resulting coupling of transition dipole moments which determine the excitonic bandstructure. We have highlighted in our paper the importance of the refractive index when comparing the optical properties of such supramolecular arrays. In particular we have shown that there is an expected reduction in the resonant coupling of neighbouring molecules, and also an overall red-shift due to adsorption on a substrate which both depend on the refractive index. The collated data in Fig. 4 confirm a systematic increase in red-shift with refractive index and it will be of interest to extend this analysis to the optical properties of other planar, flat-lying molecules, and also to alternative substrates.  This combination of AFM with the capability of molecular resolution under ambient conditions, fluorescence microscopy, and the solution deposition of supramolecular arrays with monolayer thickness provides new insights into the influence of environment on the properties of organic molecules and, in particular, demonstrates the importance of refractive index on optical transitions, of relevance to both fundamental studies and technologically-significant organic/inorganic heterostructures.

**Acknowledgments**

This work was supported by the Engineering and Physical Sciences Research Council [grant numbers EP/N033906/1 and EP/K005138/1]; the Leverhulme Trust [grant number RPG-2016-104]. Nottingham Nanoscale and Microscale Research Centre enabled access to the Horiba LABRAM instrument. K.W. and T.T. acknowledge support from the Elemental Strategy Initiative conducted by the MEXT, Japan and the CREST (JPMJCR15F3), JST. EB acknowledges the financial support of an ERC Consolidator grant.

.

**Methods**

Substrates are prepared by mechanically exfoliating hBN flakes from mm-scale crystals using the scotch tape method. hBN flakes are deposited from a loaded tape onto thermally oxidised silicon wafers, with an oxide thickness of 300 nm, and thermally deposited chromium on silicon dioxide. The flakes are cleaned by immersion in toluene for approximately 12 hours and annealing in $H_2$:Ar (5% : 95%) at 400 °C for 8 hours. In some cases, brief flame annealing prior to the deposition of organic molecules, as described previously, is carried out.

The PTCDI-melamine network is deposited onto clean hBN flakes from a dimethylformamide (DMF) solution of PTCDI and melamine molecules with concentrations of ~0.5 µM and 0.66mM respectively. The deposition was carried out at 100°C and the sample was subsequently washed with 1ml of DMF and dried in $N_2$-stream for ~1min. PTCDI is formed by rinsing samples of the pre-formed PTCDI-melamine network with ~100ml of ultra-pure water, in order to remove the soluble melamine species and leave insoluble PTCDI on the surface. PTCDA is deposited from 0.03mM ethanolic solution for 25 hours at room temperature. The sample was dried in $N_2$-stream afterwards.

Fluorescence spectroscopy is carried out using a Horiba LabRAM HR spectrometer, equipped with a 532 nm excitation laser. Laser powers in the range of 1-50 µW are used to reduce photo-bleaching and damage to the sample. The sample morphology is determined using AFM, carried out under ambient conditions in tapping mode using the Asylum Research Cypher S with Mulit75Al-G silicon cantilevers from Budget Sensors.

Fusti-Molnar, A. Ghysels, A. Golubeva-Zadorozhnaya, J. Gomes, M.W.D. Hanson-Heine, P.H.P. Harbach, A.W. Hauser, E.G. Hohenstein, Z.C. Holden, T.C. Jagau, H. Ji, B. Kaduk, K. Khistyaev, J. Kim, J. Kim, R.A. King, P. Klunzinger, D. Kosenkov, T. Kowalczyk, C.M. Krauter, K.U. Lao, A.D. Laurent, K. V. Lawler, S. V. Levchenko, C.Y. Lin, F. Liu, E. Livshits, R.C. Lochan, A. Luenser, P. Manohar, S.F. Manzer, S.P. Mao, N. Mardirossian, A. V. Marenich, S.A. Maurer, N.J. Mayhall, E. Neuscamman, C.M. Oana, R. Olivares-Amaya, D.P. Oneill, J.A. Parkhill, T.M. Perrine, R. Peverati, A. Prociuk, D.R. Rehn, E. Rosta, N.J. Russ, S.M. Sharada, S. Sharma, D.W. Small, A. Sodt, T. Stein, D. Stück, Y.C. Su, A.J.W. Thom, T. Tsuchimochi, V. Vanovschi, L. Vogt, O. Vydrov, T. Wang, M.A. Watson, J. Wenzel, A. White, C.F. Williams, J. Yang, S. Yeganeh, S.R. Yost, Z.Q. You, I.Y. Zhang, X. Zhang, Y. Zhao, B.R. Brooks, G.K.L. Chan, D.M. Chipman, C.J. Cramer, W.A. Goddard, M.S. Gordon, W.J. Hehre, A. Klamt, H.F. Schaefer, M.W. Schmidt, C.D. Sherrill, D.G. Truhlar, A. Warshel, X. Xu, A. Aspuru-Guzik, R. Baer, A.T. Bell, N.A. Besley, J. Da Chai, A. Dreuw, B.D. Dunietz, T.R. Furlani, S.R. Gwaltney, C.P. Hsu, Y. Jung, J. Kong, D.S. Lambrecht, W. Liang, C. Ochsenfeld, V.A. Rassolov, L. V. Slipchenko, J.E. Subotnik, T. Van Voorhis, J.M. Herbert, A.I. Krylov, P.M.W. Gill, and M. Head-Gordon, Mol. Phys. **113**, 184 (2015).